# A Lossless Data Hiding Technique based on AES-DWT

Francisco Rubén Castillo Soria[1], Gustavo Fernández Torres[2], Ignacio Algredo-Badillo[1]

[1] Computer Engineering, University of the Istmo
Santo Domingo Tehuantepec, Oaxaca, 70760, Mexico

[2] Petroleum Engineering, University of the Istmo
Santo Domingo Tehuantepec, Oaxaca, 70760, Mexico

**Abstract**
In this paper we propose a new data hiding technique. The new technique uses steganography and cryptography on images with a size of 256x256 pixels and an 8-bit grayscale format. There are design restrictions such as a fixed-size cover image, and reconstruction without error of the hidden image. The steganography technique uses a Haar-DWT (Discrete Wavelet Transform) with hard thresholding and LSB (Less Significant Bit) technique on the cover image. The algorithms used for compressing and ciphering the secret image are lossless JPG and AES, respectively. The proposed technique is used to generate a stego image which provides a double type of security that is robust against attacks. Results are reported for different thresholds levels in terms of PSNR.

## 1. Introduction

Both steganography and cryptography are processes for information hiding, which require different implementation strategies. However, cryptography and steganography can be complementary processes, enabling a higher level of information security. Cryptography hides the information in an unintelligible way, whereas steganography hides information by using a carrier (plain message) to convey the hidden information. This paper proposes a new technique which uses a combination of these two processes, see Fig. 1. Powerful algorithms are used to create a hidden message which, if discovered, would still remain ciphered.

Steganographic techniques are classified into two categories [1], those which operate in the spatial domain and those which need the domain of a transformation.

LSB (least significant bit) is a widely-used spatial technique, which provides good results operating on high resolution images with great amounts of color. However, transmitting large images via the Internet may appear suspicious. This problem can be solved using strategies in the Wavelet-transform domain. A characteristic of this transformation is that the values of its matrix are dispersed and many of them are zeros, which are mainly distributed in the first quadrant (HH, more details in Section 2) and represent higher frequencies. If more space is required, then it is possible to use the hard-thresholding technique. This technique is based on a value $K$ that is compared with all the transform coefficients and those with absolute values under a certain threshold $K$ are set to zero. The newly transformed image will have a large quantity of zeros and will be preserved with low distortion. Using the space represented by these zeros, it is possible to insert either a secret image or file which will appear hidden when the inverse transformation has been completed.

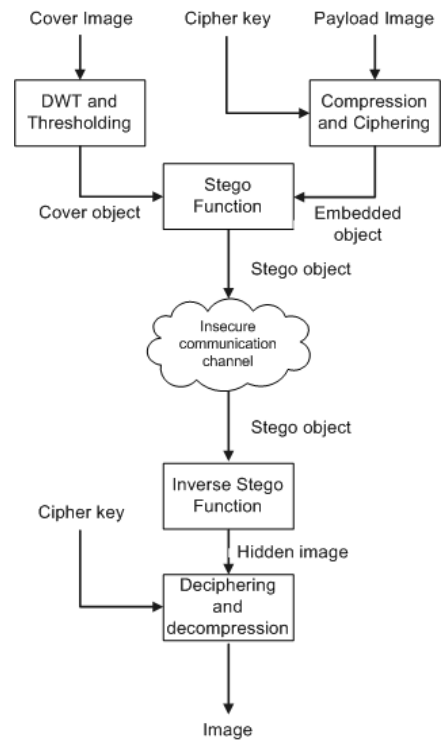

Fig. 1 Block diagram of the proposed technique

The imposed restrictions are desirable characteristics in any steganographic system: first, the size of stego image (carrier image) does not increase, because it may create suspicion; and second, the hidden image (secret image) can be reconstructed without error, allowing the use of cryptographic technology. The second restriction is important, because the hidden image has been compressed and ciphered, and it is necessary to recover files without errors. The algorithm used for compression is lossless JPG and for ciphering is AES.

On the receiver side, the cipher image is recovered from the carrier image. An algorithm runs through the transformed stego image and detects sequences with absolute values smaller than $K$. These values are arranged to obtain the cipher image.

The cryptographic algorithm used is AES cipher [2], which is classified as a symmetric key algorithm. In this case, both the transmitter and receiver know the unique key to cipher and decipher the message, hence the name of symmetric key. AES is based on a substitution-permutation network and operates in a finite field (see Fig. 2), which works on fixed-size blocks from the message. This algorithm requires little time, needs low memory and provides a high-level of security.

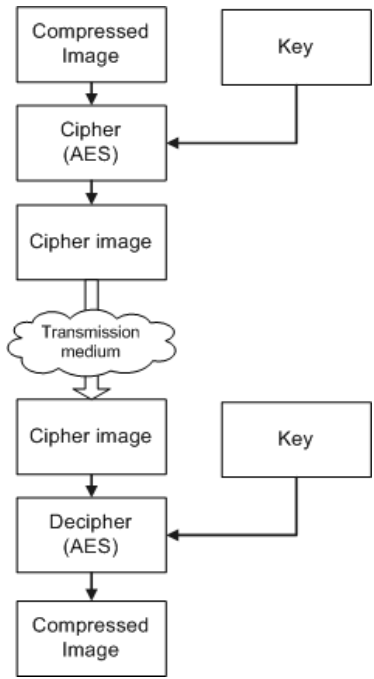

Fig. 2 Ciphering and deciphering processes using AES algorithm

The rest of this paper is organized as follows. Section 2 describes Haar transform. Section 3 discusses the proposed techniques in details. Section 4 provides analysis and results for the scheme based on cryptography and steganography. Finally, Section 5 presents conclusions.

## 2. The Haar-DWT

There are many recent publications where the Haar transform is used for hiding images. In [3], the reported PSNR range reaches 51 dB, however there is no discussion of rounding and enlargement of the stego image. In [4], the PSNR range varies from 19 to 25 dB, so that the signal presents considerable distortion. In [5], the results of the technique are satisfactory but it requires an extra recovery matrix, which is not desirable because it implies extra processing.

To describe the Haar-wavelet, definitions will first be introduced to clarify the concept of a wavelet transform [6][7].

The classes of functions represented by a wavelet transform are those that are square integrable on the real line. This class is denoted as $L_2(\Box)$. Thus, the notation $f(x) \in L_2(\Box)$ means

$$\int_{-\infty}^{\infty} |f(x)|^2 \, dx < \infty. \quad (1)$$

A function $\psi(x)$ is an orthogonal wavelet if the set $\{\psi_{j,k}\}_{j,k \in \Box}$ of functions defined by

$$\psi_{j,k}(x) = 2^{j/2} \psi(2^j x - k) \quad (2)$$

where $-\infty < j, k < \infty$ are integers, forms an orthonormal basis of $L_2(\Box)$ [6]. The integer $j$ determines a dilatation named binary dilatation by $2^j$, while $k$ determines a translation named dyadic translation by $k/2^j$.

The preceding wavelet set forms an orthonormal basis if, first,

$$\langle \psi_{j,k}, \psi_{l,m} \rangle = \delta_{j,l} \delta_{k,m}, \quad (3)$$

where $l$ and $m$ are integers, $\delta_{j,k}$ is the Kronecker delta function, and $\langle \cdot, \cdot \rangle$ indicates the inner product; and second,

if any function $f(x) \in L_2(\mathbb{R})$ can be written as

$$f(x) = \sum_{j=-\infty}^{\infty} \sum_{k=-\infty}^{\infty} c_{j,k} \psi_{j,k}(x) \quad (4)$$

where the transform coefficients are again given by inner products; that is

$$c_{j,k} = \langle f(x), \psi_{j,k}(x) \rangle = \\ = 2^{j/2} \int_{-\infty}^{\infty} f(x) \psi(2^j x - k) dx \quad (5)$$

Equations (4) and (5) specify a wavelet series expansion of $f(x)$ relative to the wavelet $\psi(x)$.

If we further restrict $f(x)$ and the basic wavelet $\psi(x)$ to functions that are zero outside the interval $[0,1]$, then the family of orthonormal basis functions can be specified by a single index $n$; that is, if we want $\psi$ take values in $[0,1]$ we must have $2^j x - k \in [0,1]$, that is, $\frac{k}{2^j} \le x \le \frac{k+1}{2^j}$, with $j = 0, 1, \ldots$. Since the values of $x$ must be in $[0,1]$, then $k$ must satisfy $0 \le k \le 2^j - 1$.

With this we define

$$\psi_n(x) = 2^{j/2} \psi(2^j x - k) \quad (6)$$

where $j$ and $k$ are functions of $n$ as follows $n = 2^j + k$ for $j = 0, 1, \ldots,$ and $k = 0, 1, \ldots, 2^j - 1$, where we consider $\psi_0(x) = 1$. The functions $\psi_n$ are the same type as the originals but restricted to the interval $[0,1]$.

In this way, since the Haar system is defined by

$$\psi = \chi_{[0,0.5)} - \chi_{[0.5,1)} = \begin{cases} 1, x \in [0,0.5) \\ -1, x \in [0.5,1), \\ 0, \text{other case} \end{cases}$$

and includes $\psi_0(x) = 1$ in the representation (6), we change the index $k$ for $k-1$, thus

$$\psi_0(x) = \psi_{00}(x) = 1,$$

$$\psi_n(x) = \psi_{j,k}(x) = \begin{cases} 2^{j/2}, & \frac{k-1}{2^j} \le x < \frac{k-\frac{1}{2}}{2^j} \\ -2^{j/2}, & \frac{k-\frac{1}{2}}{2^j} \le x < \frac{k}{2^j} \\ 0, & \text{other case} \end{cases}$$

with $n = 2^j + k - 1$ for $j = 0, 1, \ldots, \quad k = 0, 1, \ldots, 2^j$.

In an analogous way, if for any $n$, $j$ is the largest integer such that $2^j \le n$, then we have that $k - 1 = n - 2^j$, is the reminder.

Now, the transform and the inverse transform are

$$f(x) = \sum_{n=0}^{\infty} c_n \psi_n(x) \quad (7)$$

and the transform coefficients are given by the inner product

$$c_n = \langle f(x), \psi_n(x) \rangle = 2^{j/2} \int_{-\infty}^{\infty} f(x) \psi_n(x) dx. \quad (8)$$

Here a continuous function is being represented by a single infinite sequence as with the Fourier series. If $f(i\Delta t)$ is a discrete function sampled with $N$ points, where $N = 2^n$, and if $\psi(x)$ is a compact dyadic wavelet, as Haar wavelet, then we can compute a discrete wavelet transform using discrete versions of (7) and (8).

Since $N = 2^n$, then $\Delta t = \frac{1}{N} = \frac{1}{2^n}$, and the values of $f(x)$ are $f\left(\frac{i}{2^n}\right), i = 0, 1, \ldots, 2^n - 1$.

The Haar transform, using the above, are defined by

$$h_0(x) = \frac{1}{\sqrt{N}},$$

$$h_n(x) = \frac{1}{\sqrt{N}} \begin{cases} 2^{j/2}, & \frac{k-1}{2^j} \le x < \frac{k-\frac{1}{2}}{2^j} \\ -2^{j/2}, & \frac{k-\frac{1}{2}}{2^j} \le x < \frac{k}{2^j} \\ 0, & \text{other case} \end{cases}$$

With the above definition, we represent the associated matrix to $h_n$ by $\mathbf{H}$, a matrix $N \times N$. Since the Haar transform is separable we have that $\mathbf{H}^{-1} = \mathbf{H}^t$ and since $\mathbf{H}$ is not symmetric, $\mathbf{H}^t$ cannot be replaced by $\mathbf{H}$, then, if $\mathbf{F}$ is an $N \times N$ image matrix, and $\mathbf{T}$ is the resulting $N \times N$ transform we have

$$\mathbf{T} = \mathbf{HFH}^t.$$

In the same way, applying $\mathbf{H}^t$ in left side of previous result and $\mathbf{H}$ on the right side we have

$$\mathbf{F} = \mathbf{H}^t\mathbf{TH}.$$

Figure 3 presents the concept of levels in the DWT. Most of zeros are in HH, LH and HL zones.

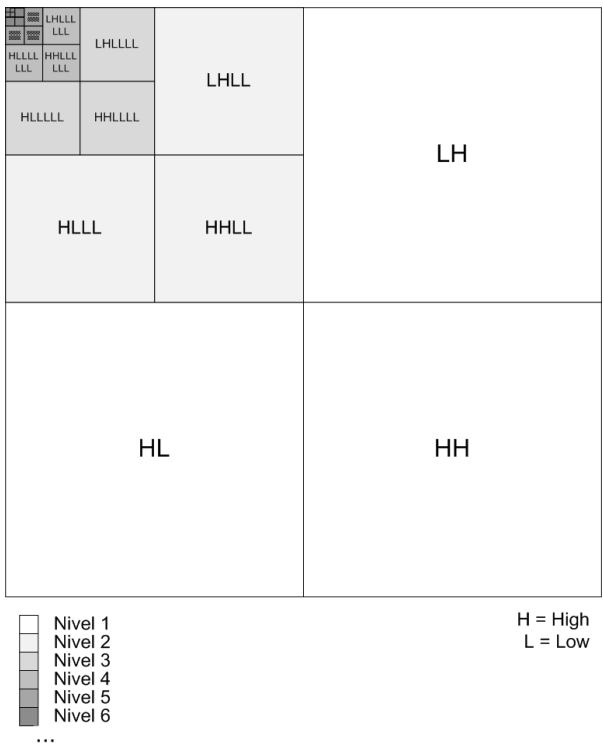

Fig. 3 The 2D-DWT

## 3. Proposed Technique

In general, the proposed technique is constituted by two main parts: i) The processing on the cover image and ii) The processing on the secret image. The complete process is described by the following steps:

1. Compute DWT of the original (cover) image,
2. Insert zeros by using hard thresholding in the DWT,
3. Choose K. The capacity or payload,
4. Define a satisfactory coding, and the bits per symbol (bps).
5. Compress and cipher the secret image (or payload).
6. Insert the secret image in the zero values of the DWT using LSB.
7. Compute the inverse DWT, and
8. Normalize and round the image (stego image).

### 3.1. Error measurement

For purposes of error measurement, the PSNR (Peak Signal to Noise Ratio) is used [6]. To define the PSNR, it is necessary to formulate the mse (mean square error).

The mse, for two monochromatic images $Y$ and $Z$ with size *MxN*, is defined as:

$$mse = \frac{1}{MN}\sum_{i=0}^{M-1}\sum_{j=0}^{N-1} \|Y(i,j) - Z(i,j)\|^2$$

So, the PSNR is defined as:

$$PSNR = 10\log_{10}\left(\frac{MAX_I^2}{mse}\right) =$$

$$= 20\log_{10}\left(\frac{MAX_I}{\sqrt{mse}}\right)$$

where $MAX_I$ is the maximum value for a pixel in the image. If the pixels are represented by $B$ bits per sample, then $MAX_I = 2B - 1$.

Figure 4 shows an original (cover) image of *Lena*. This image has a size of 256x256 in 8-bit grayscale.

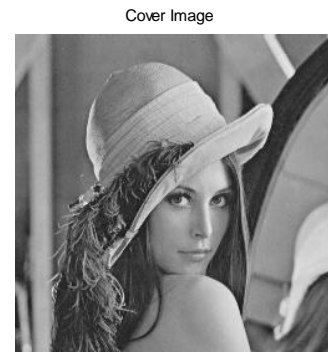

Fig. 4 Cover Image

In step 1, the Haar-wavelet transform of the cover image is computed (see Fig. 5). This transform shows that a great amount of coefficients has a value close to zero (black pixels).

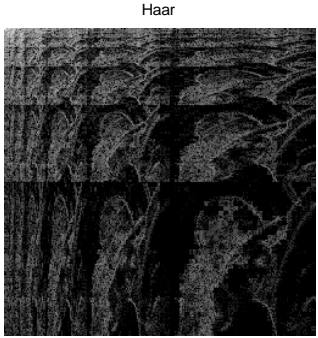

Fig. 5 Haar coefficients

In step 2, the hard thresholding technique is applied on the Haar coefficients. This technique is based on replacing certain coefficients, which are in a range defined by *K*.

$$\tilde{s}_i = \begin{cases} 0, & |s_i| < k \\ s_i, & |s_i| \geq k \end{cases}$$

where $s_i$ is the *i-th* transform coefficient, obtaining new transformed data.

In the proposed technique, this can be seen as enabling larger space for hiding an image with minimal error.

The more the value of *K* is increased, the more space for hiding data is obtained. In addition more noise is introduced in the system.

For step 3, it is important to determine an adequate coding in order to obtain the hidden image without errors. To achieve this, the error dispersion generated by the normalization and rounding should be computed.

On one hand, the normalization process is necessary because the inverse transform has components out of the valid range. On the other hand, rounding is necessary for obtaining an 8-bit quantization.

To compute the dispersion error a test payload file is used in the process.

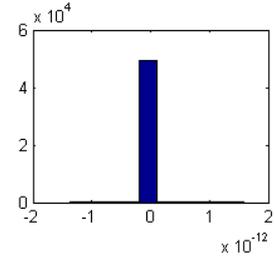

(a)

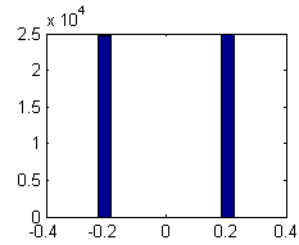

(b)

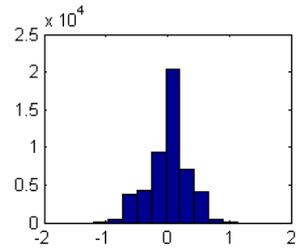

(c)

Fig. 6 The error dispersion for the reconstruction (a) without adjustment. (b) by using normalization and (c) by using rounding

If the cover image is processed without making any adjustments, the error dispersion is low (see Fig. 6a). However, when normalization and rounding are applied, the error is increased in a considerable form, which has an undesirable impact on the image reconstruction (see Fig. 6b and Fig. 6c).

The total dispersion error takes into account both normalization and rounding effects. Figure 7 shows that the error is concentrated for values less than ±1.5. Considering the previous error analysis in step 4, the symbols must have a minimum distance of 3. In step 5, a coder is designed which takes into consideration the capacity requirement within the tolerable error.

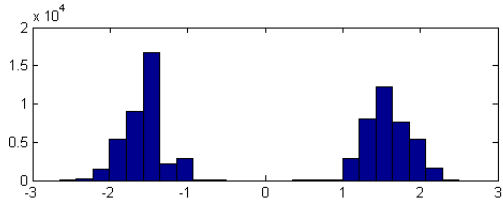

Fig 7. Total dispersion error

Step 6 only operates on the secret image. The secret image is compressed and then it is ciphered.

For compression, the lossless JPEG algorithm is used, whereas, the AES algorithm is computed for the ciphering. The compressed image is fed into the cipher algorithm generating a cipher image, see Fig. 8.

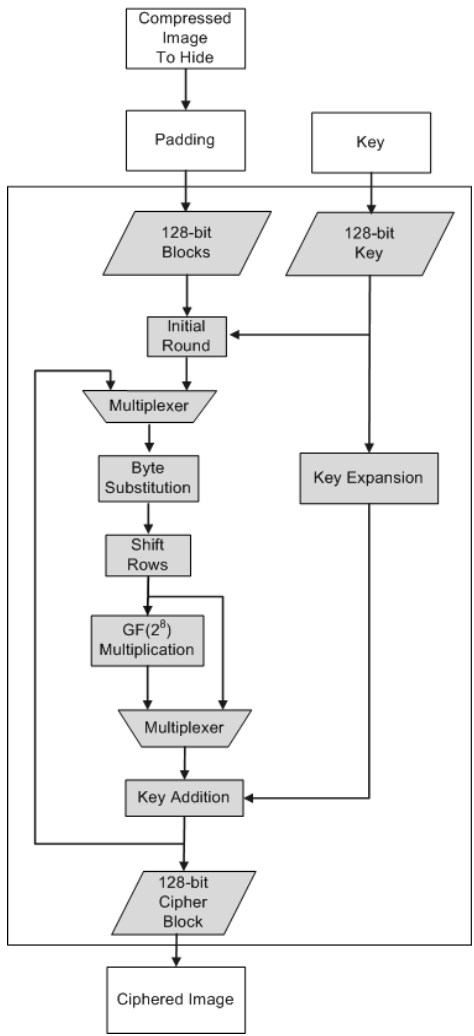

Fig. 8 Block diagram for computing AES cipher algorithm

Steps 7 and 8 are described for the cases of 1b/s, 2b/s and 3b/s (bits per symbol). The simplest case is 1b/s. In this case, the decision threshold is placed at zero. Table 1 shows the used coding for 1b/s 2b/s and 3b/s. The minimum distance is 3.

Table 1. Coding

| Bits per symbol | Transmitted bits | Symbol's amplitude |
|---|---|---|
| 1 b/s | 0 | -1.5 |
|  | 1 | 1.5 |
| 2 b/s | 00 | -4.5 |
|  | 01 | -1.5 |
|  | 10 | 1.5 |
|  | 11 | 4.5 |
| 3 b/s | 000 | -10.5 |
|  | 001 | -7.5 |
|  | 010 | -4.5 |
|  | 011 | -1.5 |
|  | 100 | 1.5 |
|  | 101 | 4.5 |
|  | 110 | 7.5 |
|  | 111 | 10.5 |

Table 2 shows the capacity of the cover image (number of bits that can be embedded into the particular cover image), the PSNR and the mse for 1b/s, 2b/s and 3b/s. These values are associated with different thresholds (K).

Table 2. PSNR and capacity

| Bits Per Symbol | K | Hiding Cap. (KB) | mse | PSNR (dB) |
|---|---|---|---|---|
| 1 b/s | 1.5 | 3.31 | 13.6 | 36.7 |
|  | 5 | 5.1 | 16.5 | 35.9 |
|  | 10 | 6.3 | 22.3 | 34.6 |
|  | 15 | 6.7 | 46.5 | 31.4 |
|  | 20 | 7.1 | 54.4 | 30.7 |
|  | 25 | 7.3 | 63.8 | 30 |
|  | 30 | 7.5 | 76.8 | 29.2 |
| 2 b/s | 5 | 10.3 | 23 | 34.4 |
|  | 10 | 12.6 | 30.4 | 33.3 |
|  | 15 | 13.5 | 54.7 | 30.7 |
|  | 20 | 14.2 | 65.6 | 29.9 |
|  | 25 | 14.6 | 73.9 | 29.4 |
|  | 30 | 15 | 86.1 | 28.7 |
| 3 b/s | 10.5 | 18.8 | 53.4 | 30.8 |
|  | 15 | 20.1 | 66.9 | 29.8 |
|  | 20 | 21.3 | 78.7 | 29.1 |
|  | 25 | 22 | 93.2 | 28.4 |
|  | 30 | 22.5 | 112 | 27.6 |

Figure 9 presents comparison cases between the three types of coding.

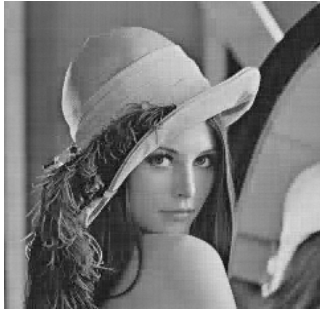

(a)

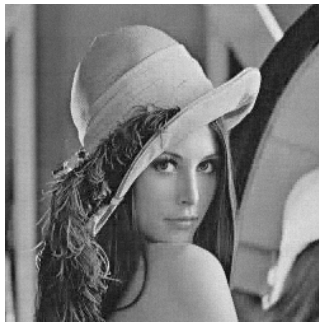

(b)

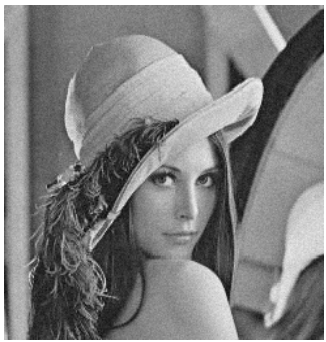

(c)

Figure 9.Stego images comparison. a) 1b/s, K=20, b) 2b/s, K=15, c) 3 b/s, K=10.5

## 4. Results

Figures 10 to 12 show some obtained results in terms of the capacity, mse and *K*.

If 1b/s is used, the cover image has higher PSNR rates. At the same time, however, it has a low capacity. In the inverse case, when 3b/s are used, the capacity for hiding is higher, but the stego image has greater distortion. In this case, the cover image is able to hide an image or file that is up to 39% of the cover image size. When a 2b/s codification is used, we have no visual evidence of tamper and a capacity of hiding up to 25%. These results can compete with previous results reported for commercial software in [8].

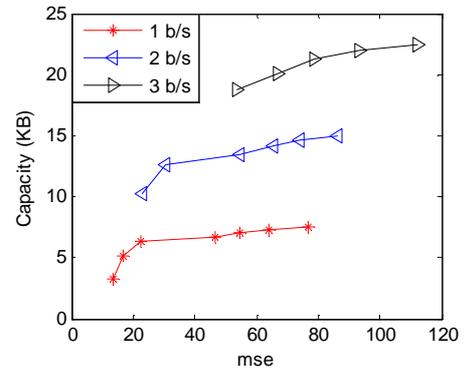

Fig 10. mse versus capacity

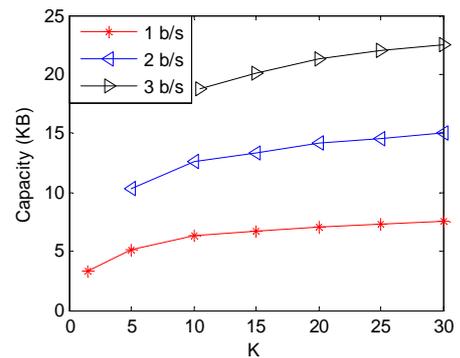

Fig 11. Thresholding *K* versus capacity

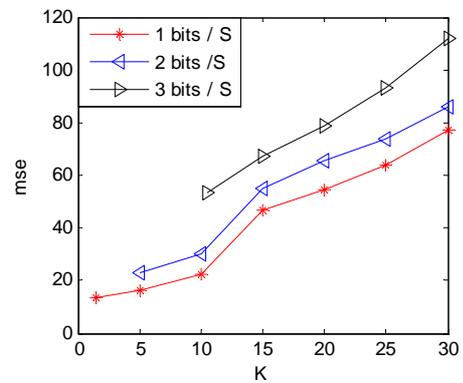

Fig 12. Thresholding *K* versus mse

**Example**

In this example three images which have a size of 256x256 pixels are used. In this case a coding of 2b/s and a threshold $K = 20$ are used. The secret image (Figure 14) was recovered without error.

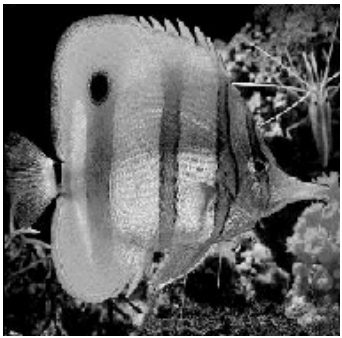

Fig. 13 Cover Image

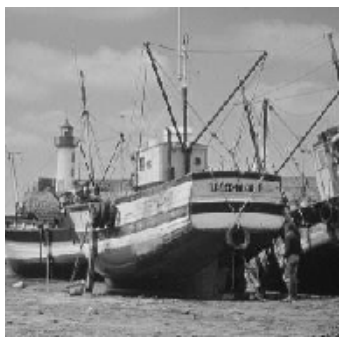

Fig. 14 Secret Image

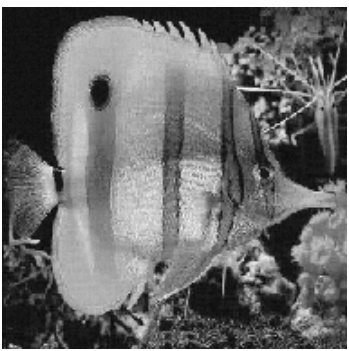

Fig. 15 Stego Image

Future work will be focused on the design of efficient hardware architecture that will speed up implementation. The main idea is design the customization of reconfigurable hardware on FPGAs (field programmable gate arrays) for this specialized application which consumes a large amount of computational resources (time and memory).

## 5. Conclusions

Results show that the proposed technique is effective for hiding images with files which are recovered without errors. The capacity depends on the quantity of bits per symbol used and K. Additionally, security was increased by using cryptography techniques, which provide double security.

**Acknowledgments**

The authors wish to acknowledge the valuable participation of professor Ted Fondulas and Alex Yealland in the proofreading of this paper. Moreover, thanks to PROMEP project 103.5/11/5266 for providing financial support.

**Francisco Ruben Castillo Soria** received his Master of Science in Telecommunications Engineering from IPN-ESIME, Research and Graduate Section in Mexico City, 2004. From 2001 to 2004 he was the Telecommunications and Computer Science Engineering Major Director of the Technological University of Baja California in B.C.S. Since 2005 he has been Research-Professor at the Istmo University, Oaxaca. His current research interests lie in simulation of telecommunications systems, protocols, multiple access and signal processing.

**Gustavo Fernández-Torres** received the B.S in Physics and Mathematics from Superior School of Physics and Mathematics of Polytechnic National Institute (ESFM-IPN) and the Master of Science degree in Pure Mathematics from Investigation Center in Mathematics (CIMAT), 2004. Since 2007, he has been Research-Professor of Petroleum Engineering Department at the Istmo University. His current research interests lie in Numerical and Functional Analysis and Operators Theory.

**Ignacio Algredo-Badillo** received the B.Eng in Electronic Engineering from Technologic Institute of Puebla (ITP) in 2002 and the M.Sc and Ph.D degrees in Computer Science from National Institute for Astrophysics, Optics and Electronics (INAOE) in 2004 and 2008, respectively. Since 2009, he has been professor of Computer Engineering at University of Istmo. He has involved in the design and development of digital systems, reconfigurable architectures, software radio platforms, cryptographic systems, FPGAs implementations, microcontrollers-based systems and hardware acceleration for specific applications.